\documentclass[aps,showpacs,prl,twocolumn]{revtex4-1}  

\usepackage{amsmath}
\usepackage{mathrsfs}
\usepackage{amssymb}
\usepackage{graphicx,epstopdf,epsfig}
\usepackage{bm}
\usepackage{color}
\usepackage{times}
\usepackage{hyperref}

\newcommand{\mr}[1]{{\mathrm{#1}}} 		
\newcommand{\mrb}[1]{{\bm{\mathrm{#1}}}} 	
\newcommand{\bra}[1]{{\langle #1 \vert}} 		
\newcommand{\ket}[1]{{\vert #1 \rangle}} 		

\newcommand{\ii}{\,\mr{i}}					
\newcommand{\Tr}{\,\mr{Tr}}				
\newcommand{\N}{\,{\mathcal{N}}}			
\newcommand{\e}{\mr{e}}					

\newcommand{\mean}[1]{\,\langle #1\rangle}					

\newcommand{\Omp}{\,\Omega_\mr{p}}		
\newcommand{\Omc}{\,\Omega_\mr{c}}		
\newcommand{\Ge}{\,\Gamma_e}			

\definecolor{clr}{rgb}{0,0.6,0.6}

\begin{document}

\title{Correlated photon emission from multi--atom Rydberg dark states}

\author{J. D. Pritchard}
\author{C. S. Adams}
\email{c.s.adams@durham.ac.uk}
\affiliation{Department of Physics, Durham University, Rochester Building, South Road, Durham DH1 3LE, UK}
\author{K. M\o{}lmer}
\email{moelmer@phys.au.dk}
\affiliation{Lundbeck Foundation Theoretical Center for Quantum System Research, Department of Physics and Astronomy, University of Aarhus, DK-8000 \r{A}rhus C, Denmark}

\date{\today}

\begin{abstract}
We consider three level atoms driven by two resonant light fields in a ladder scheme where the upper level is a highly excited Rydberg state. We show that the dipole--dipole interactions between Rydberg excited atoms prevents the formation of single particle dark states and leads to strongly correlated photon emission from atoms separated by distances large compared to the emission wavelength. For two atoms, correlated photon pairs are emitted with an angular distribution given by a coherent sum of the independent dipolar fields.
\end{abstract}

\maketitle

The cooperative emission of light, as in superradiance \cite{dicke54,gross82}, offers an interesting paradigm for photon entanglement and non-classical light generation. Cooperative effects for few particle systems have been observed for both trapped ions \cite{devoe96} and molecules \cite{hettich02}, and also quantum dots \cite{scheibner07}. For cooperative effects to dominate,  the emitters should not be resolved by the radiation field so that interference effects can occur, or the resonant dipole--dipole interaction should be large enough to form correlated atomic states, i.e., it should typically exceed the natural linewidth of the transition. In practice, this imposes stringent demands on the spatial distributions of emitters, and for the dipole interaction to be effective, the emitters should be separated by much less than the optical wavelength \cite{agarwal74,beige98}.

In this Letter we show that coupling the dipoles associated with transitions between highly excited Rydberg states to an optical atomic transition leads to cooperative emission with considerably less stringent demands on the spatial confinement of the emitters.
The optical and long wavelength dipoles become coupled by electromagnetically induced transparency (EIT) \cite{boller91} involving highly excited Rydberg states \cite{mohapatra07}. The effect of the strong dipole--dipole interactions between Rydberg states is to modify the EIT dark state such that only a single atom within the Rydberg interaction range can contribute to the EIT \cite{pritchard10}, whilst the remaining atoms scatter light like resonant two-level systems. This effect is a manifestation of the dipole blockade mechanism where the dipole induced level shifts prevent multiple Rydberg excitation \cite{lukin01} within a volume $\textstyle{4\over 3}\pi R_{\rm b}^3$, where the blockade radius $R_{\rm b}$ is typically of the order of a few microns. This blockade effect leads directly to entangled atomic superposition states that may be exploited for applications in quantum information processing \cite{saffman10}. In addition, the blockade mechanism may be used to modify light transmission \cite{pritchard10, ates11,sevincli11} giving rise to non--classical states of light \cite{honer11} and strong photon--photon interactions \cite{shahmoon10,gorshkov11,petrosyan11}. Blockaded superpositions can also be mapped into exotic states of light \cite{olmos10,pohl10,nielsen10}.

\begin{figure}[t]
\includegraphics{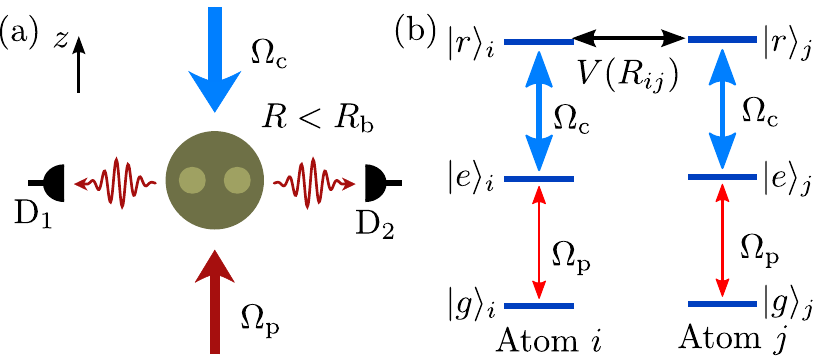}
\caption{(Color online). (a) For atoms confined within $R<R_\mathrm{b}$, dipole blockade modifies the EIT dark state leading to correlated photon emission from the blockade region which can be measured using detectors D$_1$ and D$_2$. (b) Level-scheme for two-atoms with a dipole-dipole coupling $V(R)$ between the Rydberg states, where $R$ is the interatomic separation.}
\label{fig1}
\end{figure}

For a few-atom system localised within a single blockade volume, shown schematically in Fig.~\ref{fig1}~(a), we show that dipole blockade leads directly to strongly correlated photon-emission for atoms separated by several microns.  This parameter regime is compatible with the current experimental setups used to demonstrate entanglement \cite{wilk10} and quantum gates \cite{isenhower10} using blockade for a pair of atoms.

Consider an ensemble of three-level atoms composed of a ground state $\ket{g}$, excited state $\ket{e}$ and Rydberg state $\ket{r}$, shown schematically in Fig.~\ref{fig1}~(b). The atoms are coupled by counter-propagating probe and coupling fields aligned parallel to the $z$-axis with wavevectors of $\mrb{k}_p=k\hat{\mrb{z}}$ and $\mrb{k}_c=-k'\hat{\mrb{z}}$ respectively. The weak probe beam resonantly drives the transition from $\ket{g}\rightarrow\ket{e}$ with Rabi frequency $\Omp$, whilst the strong coupling laser with Rabi frequency $\Omc$ is resonant with the transition from $\ket{e}\rightarrow\ket{r}$. The Hamiltonian for the system is given by $\mathscr{H}=\mathscr{H}_{al}+\mathscr{H}_\mr{dd}$, where  $\mathscr{H}_\mr{al}=-\hbar\sum_i^\mathcal{N}[\Omp\hat{\sigma}_{eg}^i\e^{\ii\mrb{k}_p\cdot\mrb{r}_i}+\Omc\hat{\sigma}_{re}^i\e^{\ii\mrb{k}_c\cdot\mrb{r}_i}+\mr{H.C.}]$ accounts for the atom-light coupling and $\mathscr{H}_\mr{dd}=-\hbar\sum_{i<j}^\mathcal{N}\hat{\sigma}_{rr}^iV(\vert \bm{r}_i-\bm{r}_j\vert)\hat{\sigma}_{rr}^i$ represents the dipole-dipole coupling between atoms $i$ and $j$ at positions $\bm{r}_i$ and $\bm{r}_j$ respectively, where $\hat{\sigma}^i_{mn}$ is the outer product operator $\ket{m}\bra{n}$ of atom $i$. In the limit of long-range Van der Waals interactions between the Rydberg states, $V(R)=-C_6/R^6$, resulting in blockade radius of $R_\mr{b}=\sqrt[6]{C_6/\Omc}$ \cite{pritchard10}. When the atoms are separated by $R>R_\mr{b}$, the atoms behave independently and the system evolves into the EIT dark state superposition of $\ket{g}$ and $\ket{r}$ resulting in perfect probe transmission as there is no loss from the short-lived state $\ket{e}$. For atoms localized within $R_\mr{b}$ however, the blockade modifies the dark state resulting in a cooperative optical non-linearity \cite{pritchard10} as only a single atom can contribute to the EIT dark state, with the remaining atoms resonantly coupling to the probe laser and scattering photons out of the beam.

The interplay between EIT and Rydberg blockade is expected to cause interesting effects in the excitation dynamics, and the purpose of this Letter is to identify its observable experimental signature in the light emitted from the atoms. In particular, we will show how the transient dynamics in and out of dark states can be unambiguously identified by the bunching and anti-bunching features in photon counting signals under steady state driving conditions. To evaluate our theoretical prediction for the photon-photon correlations in the scattered field from the atomic system we consider the normalised second-order correlation function $g^{(2)}(\tau)$ of the probe electric field. The positive-frequency component of the probe electric field operator at position $\mrb{r}$ is given by the source-field expression $\hat{\bm{E}}^{(+)}(\mrb{r},t)=\hat{\bm{E}}_\mr{f}^{(+)}(\mrb{r},t)+\hat{\bm{E}}_\mr{sc}^{(+)}(\mrb{r},t)$ \cite{loudon97}, where $\hat{\bm{E}}_\mr{f}^{(+)}(\mrb{r},t)$ is the incident probe field and $\hat{\bm{E}}_\mr{sc}^{(+)}(\mrb{r},t)$ is the radiation field of the atomic dipole. For an ensemble of $\mathcal{N}$-atoms located at positions $\mrb{r}_i$, the source-field term in the far field ($k\vert\mrb{r}-\mrb{r}_i\vert\gg1$ for all $i$) is given by \cite{agarwal74}
\begin{equation}\label{eq:field}
\hat{\bm{E}}^{(+)}_\mr{sc}(\mrb{r},t) \!=\! - \frac{k^2(\mrb{d}_\mr{eg}\!\times\!\hat{\mrb{r}})\!\times\!\hat{\mrb{r}}}{4\pi\varepsilon_0 r}\!\displaystyle\sum_i^{\N{}}\e^{-\mr{i}k\hat{\mrb{r}}\cdot\mrb{r}_i}\hat{\pi}_i^-\!\!\left(t\!-\!r/c\right),
\end{equation}
where $\mrb{d}_{eg}$ is the dipole moment and $\hat{\pi}^-_i=\ket{g}\bra{e}$ is the lowering operator for the atomic dipole of the $i$th atom. If we only consider detector positions off-axis with respect to the probe laser, the incident field $\hat{\bm{E}}_\mr{f}^{(+)}(\mrb{r},t)$ vanishes, and the electric field reduces to the sum (\ref{eq:field}) over the dipole operators for the system. Absorbing the geometric factors into the function $f(\mrb{r})$, the scattered electric field is $\hat{\bm{E}}^{(\pm)}(\mrb{r},t) = f(\mrb{r})\hat{\Pi}^\mp\left(\mrb{r},t-r/c\right)$, where $\hat{\Pi}^\pm$ are the combined raising and lowering operators for the ensemble,
\begin{equation}\label{eq:PI}
\hat{\Pi}^\pm(\mrb{r},t) = \displaystyle\sum_i^{\N{}}\e^{\pm\mr{i}k\hat{\mrb{r}}\cdot\mrb{r}_i}\hat{\pi}_i^\pm(t).
\end{equation}

For a pair of detectors $D_1$ and $D_2$ shown in Fig.~\ref{fig1}~(a) at positions $\mrb{r}_A$ and $\mrb{r}_B$, the second order mutual correlation reduces to
\begin{equation}\label{eq:g2}
g^{(2)}(\tau) \!=\! \frac{\langle\hat{\Pi}^+(\mrb{r}_A,t)\hat{\Pi}^+(\mrb{r}_B,t\!+\!\tau)\hat{\Pi}^-(\mrb{r}_B,t+\tau)\hat{\Pi}^-(\mrb{r}_A,t)\rangle}{\langle\hat{\Pi}^+(\mrb{r}_A,t)\hat{\Pi}^-(\mrb{r}_A,t)\rangle\langle\hat{\Pi}^+(\mrb{r}_B,t\!+\!\tau)\hat{\Pi}^-(\mrb{r}_B,t\!+\!\tau)\rangle},
\end{equation}
where $\mean{\ldots}$ denotes a trace over the density matrix $\sigma$ for the atomic system. The unnormalized correlation function, $G^{(2)}(t,t+\tau)$, in the numerator of Eq.~\ref{eq:g2} can be evaluated using the quantum regression theorem as
\begin{equation} \label{eq:qj}
G^{(2)}(t,t+\tau) = \Tr\{\hat{\Pi}^-(\mrb{r}_B)\sigma_\mr{cond}(t;t+\tau)\hat{\Pi}^+(\mrb{r}_B)\},
\end{equation}
where $\sigma_\mr{cond}(t;t)=\hat{\Pi}^-(\mrb{r}_A)\sigma(t)\hat{\Pi}^+(\mrb{r}_A)$ is the conditional density matrix for the system following the detection of a photon on detector $D_1$ at time $t$ and application of the corresponding quantum jump to the atomic state \cite{molmer96}. The subsequent time evolution of the conditional density matrix $\sigma_\mr{cond}(t;t+\tau)$ with respect to $\tau$ is calculated using the same master equation as for the normal density matrix, $\dot{\sigma}=\ii/\hbar[\sigma,\mathscr{H}]+\mathcal{L}(\sigma)$, where $\mathcal{L}(\sigma)$ is the Lindblad operator \cite{lindblad76} accounting for spontaneous emission from state $\ket{e}$ at rate $\Ge$
\begin{equation}
\mathcal{L}(\sigma) = \displaystyle\sum_i^\mathcal{N}-\frac{1}{2}(C_e^{i\dagger} C^i_e \sigma+\sigma C^{i\dagger}_e C^i_e)+C^i_e\sigma C_e^{i\dagger},
\end{equation}
with $C_e^i=\sqrt{\Ge}\hat{\sigma}^i_{ge}$. Note that we assume relatively small solid angles for the detectors, and the master equation treats the unobserved spontaneous emission as individual atom events due to the large atomic separation. Due to the relatively long lifetime of the Rydberg state, the spontaneous emission and other losses from $\ket{r}$ can be neglected. For the case of CW probe and coupling lasers, $\sigma(t)$ is equivalent to the steady-state of the system, leading to $g^{(2)}(\tau)=1$ at long times as the conditional density matrix evolves back to the steady-state of the system.

\begin{figure}[t!]
\includegraphics{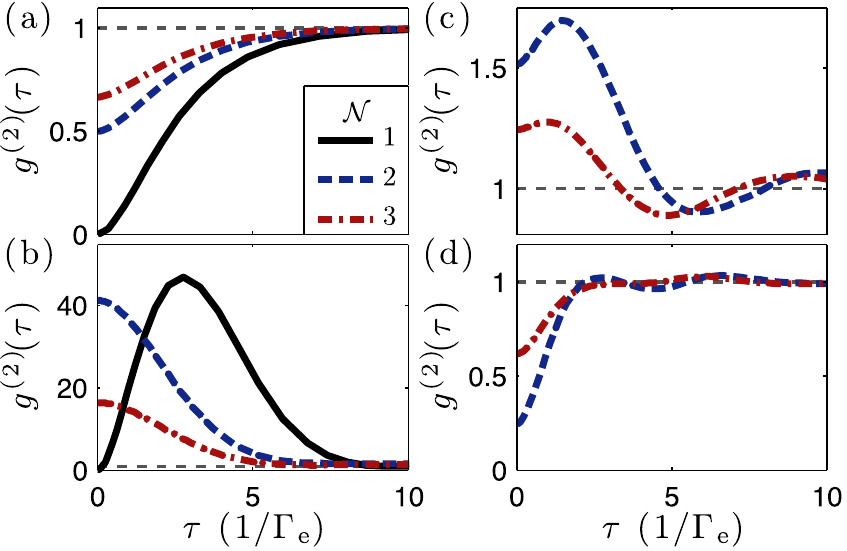}
\caption{(Color online). $\mathcal{N}$-atom fluorescence correlations. (a) Independent two-level atoms for $\Omp=\Ge/5$ displaying anti-bunching. (b)--(d) interacting EIT system with $\Omc=\Ge$, $V(r_{ij})=2\Ge$ and $\Omp=\Ge/5$, $\Ge/2$ and $\Ge$ respectively.}
\label{fig2}
\end{figure}

If the atoms move by just few metres per second, the relative phase factors in the lowering operators (\ref{eq:PI}) will change on the time scale of the atomic dynamics, and the the cross-phase terms in the observed intensity signals will vanish as if the atoms emit photons incoherently. In this case, the operators representing the detection at each detector are well approximated by an operator sum over individual atomic contributions: $\hat{\Pi}^-\sigma\hat{\Pi}^+ \rightarrow \sum_i^{\N{}} \hat{\pi}_i^-\sigma\hat{\pi}_i^+$. We calculate intensity correlation functions for this situation, and first consider the familiar example of non-interacting two-level atoms. The correlation function for a weak probe intensity of $\Omp=\Ge/5$ and $\Omc=0$ for $\mathcal{N}=1$ to 3 atoms is plotted in Fig.~\ref{fig2}~(a). This clearly illustrates anti-bunching ($g^{(2)}(0)<1$) in the resonance fluorescence of a two-level atom \cite{kimble76,kimble77} due to the finite excitation time $\tau\sim1/\Omp$ preventing a single atom from emitting one photons immediately after another. For more than one atom, the {{possibility that different atoms emit}} simultaneously reduces the visibility of the anti-bunching, and for large atom  numbers prevents the observation of the single-photon character of the scattered light.

The addition of a strong coupling laser with $\Omc=\Ge$ to a strongly blockaded ensemble with an isotropic dipole-dipole interaction $V(\vert \mrb{r}_i-\mrb{r}_j\vert)=2\Ge$ dramatically modifies the correlation function, shown in Fig.~\ref{fig2}~(b). For a single atom the curve is initially anti-bunched with $g^{(2)}(0)=0$, however this increases to give $g^{(2)}(\tau)\gg1$ at $\tau\sim1/\Omp$, resulting in very strong bunching at short times. The interpretation of this behaviour is that for a single atom the interactions play no role and the steady-state for the system is the EIT dark state. If a photon is detected at $\tau=0$, this causes a projection of the atom into the ground state with a finite component orthogonal to the dark-state. The norm of this component yields the probability for a second photon to be emitted during the subsequent evolution of the atom back into the dark state. Note, however that the photon bunching resulting from this transient atomic dynamics is, in practice, difficult to observe as the probability to emit the first photon is very small due to the EIT condition.

Adding more atoms to the system in the non-interacting limit results in similar correlation functions as the $\mathcal{N}=1$ curve due to the independent nature of the emitters. In contrast, for a blockaded ensemble, the behavior for $\N{}>1$ reveals a transition from anti-bunching to strong bunching at $\tau=0$, with photons most likely to arrive simultaneously at the two detectors. This bunching can be understood from the analytic EIT dark state for the interacting two-atom system in \cite{moller08}, that has a $\ket{ee}$ component mixed into the wavefunction in place of the blockaded $\ket{rr}$ state. This component decays by the emission of two photons within a few spontaneous lifetimes. This appears as bunching in the correlation function as shown in Fig.~\ref{fig2}~(c). In (d) the correlations for $\Omp=\Ge$ is plotted, showing that for a strong probe field the blockade condition is violated and the light becomes anti-bunched at short times, similar to the correlations for the 2- and 3-atom signals for independent two-level atoms, Fig.~\ref{fig2}~(a).

\begin{figure}[b]
\includegraphics{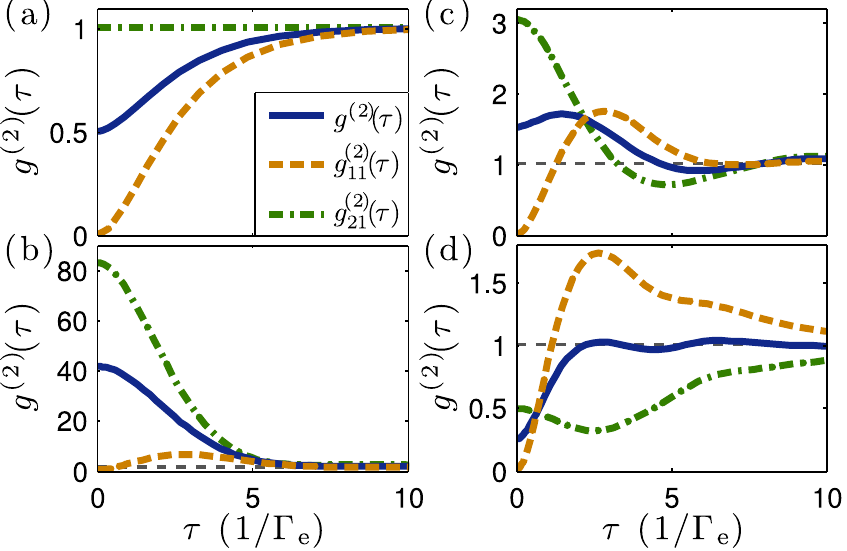}
\caption{(Color online). Self- and cross-correlations in emsission from two distinguishable atoms. (a) $\Omp=\Ge/5$. Two-level atoms have an independent cross-correlation as the atoms are non-interacting. (b)--(d) Interacting EIT system with $\Omc=\Ge$, $V(r_{ij})=2\Ge$ and $\Omp=\Ge/5$, $\Ge/2$ and $\Ge$ respectively. This shows the bunching arises from the strong cross-correlation between the atoms, which are correlated by the dipole blockade.}
\label{fig3}
\end{figure}

Due to the long-range interaction of the Rydberg dipole, typically the blockade radius $R_\mr{b}\sim5~\mu$m, 
{{spatially resolved emission is}} possible from a few-atom system. In the limit of distinguishable emission where detector $D_1$ only collects photons from atom $i$, and detector $D_2$ from atom $j$, it is possible to consider the self- and cross-correlations between atoms $i$ and $j$ defined by
\begin{equation}
g_{ij}^{(2)}(\tau) = \frac{\langle\hat{\pi}_i^+(t)\hat{\pi}_j^+(t+\tau)\hat{\pi}_j^-(t+\tau)\hat{\pi}_i^-(t)\rangle}{\langle\hat{\pi}_i^+(t)\hat{\pi}_i^-(t)\rangle\langle\hat{\pi}_j^+(t+\tau)\hat{\pi}_j^-(t+\tau)\rangle},
\end{equation}
which provide an insight into whether the emission from one atom is dependant upon the emission of a neighbouring atom.

Figure~\ref{fig3} shows the results for the case of a pair of atoms, calculated for the same parameters as above. For the probe-only system in (a), the self-correlation $g_{11}(\tau)$ is equal to the single-atom correlation function in Fig.~\ref{fig2}~(a) and the cross-correlation $g_{21}^{(2)}(\tau)$ is equal to unity for all times, showing the two atoms behave independently as expected. For the interacting EIT system however, the self- and cross-correlations show different behaviour. In the weak-probe limit (b), the bunching observed in the full correlation function arises due to the cross-correlations between the atoms, verifying the interpretation of the photons coming within a time $1/\Gamma_e$ of each other from the population of $\ket{ee}$ as discussed above. This also demonstrates that Rydberg EIT leads to a cooperative effect at the single-photon level, with the dipole emission of each atom strongly dependent upon that of its neighbour. As the probe power increases, the self-correlation in (c) and (d) is approximately constant whilst the cross-correlation switches from being bunched to anti-bunched, showing the transition from a blockaded system with cooperative emission (c) to a weakly blockaded system with suppressed emission such that one atom is less-likely to emit a photon if the other atom has emitted one already (d). These results show that it is possible to demonstrate the strong Rydberg interactions by the highly correlated fluorescence emission from a pair or a small number of atoms.

\begin{figure}[t]
	\includegraphics{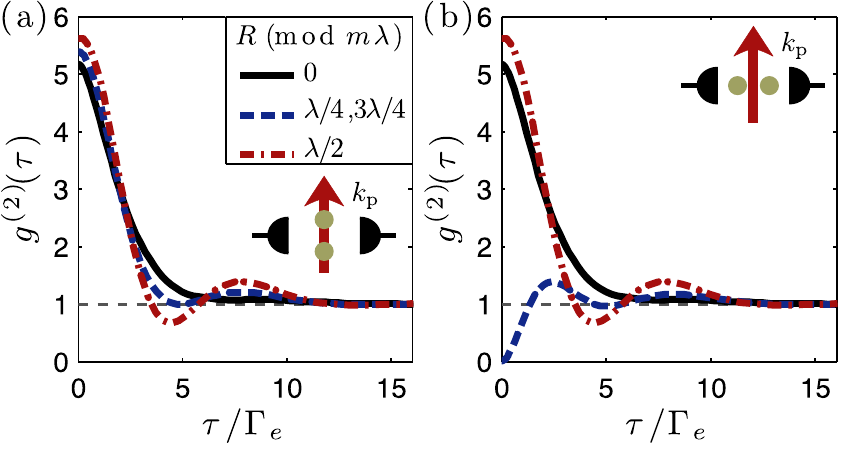}
	\caption{(Color online) Correlation function for coherent emission for different separations $R$, where $m$ is an integer. (a) Atoms aligned parallel to the probe beam are bunched for all separations. (b) The perpendicular configuration shows anti-bunching due to destructive interference for $R=m\lambda+\lambda/4,3\lambda/4$. \label{fig4}}
\end{figure}

Figure~\ref{fig4} shows the correlations calculated for coherent emission between a pair of atoms at rest with variable interatomic separation $R<R_\mr{b}$ using the position dependent phases in the dipole operator $\hat{\Pi}^\pm$ of Eq.~(\ref{eq:PI}) with detectors placed orthogonal to the probe laser, using $\Omp=\Ge/2$, $\Omc=\Ge$ and $V(r_{21})=2\Ge$. For atoms aligned parallel to the probe beam, the correlation function is bunched independent of the interatomic separation. However, emission from atoms aligned parallel to the direction towards the detectors, Fig.~\ref{fig4}(b), shows a transition from bunching to anti-bunching at $R=m\lambda+\lambda/4,3\lambda/4$, due to the density matrix conditioned by the first detection event showing destructive interference between the atomic source fields in the direction of the second detector. It is well known that the first detection events "lock" the relative phases and hence determine the direction of superradiant emission from a symmetric ensemble of atoms \cite{gross82}. In combination with EIT and Rydberg blockade, we see that this phenomenon may be observed in coincidence signals under steady state conditions.

In summary, in this Letter we have demonstrated how the dipole blockade mechanism leads to the generation of highly correlated photon emission from multi-atom Rydberg dark states. Due to the large length scales associated with the blockade radius, this reduces the experimental requirements for observing cooperative emission from atoms separated by greater than an optical wavelength. For atoms at rest, the correlated photon emission is also directionally dependent and sensitive to atomic position and may, for mesoscopic ensembles of few hundred atoms, provide interesting practical sources of non-classical light.
The second-order correlation function provides additional information about the atomic state and, as the present work illustrates, the understanding of temporal conditioned dynamics of the atoms can be confirmed by measurement of the light statistics, offering an additional probe of strongly interacting Rydberg ensembles.

\begin{acknowledgments}
JDP and CSA acknowledge support from the UK EPSRC. KM acknowledges support from the EU integrated project AQUTE.
\end{acknowledgments}


\begin{thebibliography}{29}%
\makeatletter
\providecommand \@ifxundefined [1]{%
 \@ifx{#1\undefined}
}%
\providecommand \@ifnum [1]{%
 \ifnum #1\expandafter \@firstoftwo
 \else \expandafter \@secondoftwo
 \fi
}%
\providecommand \@ifx [1]{%
 \ifx #1\expandafter \@firstoftwo
 \else \expandafter \@secondoftwo
 \fi
}%
\providecommand \natexlab [1]{#1}%
\providecommand \enquote  [1]{``#1''}%
\providecommand \bibnamefont  [1]{#1}%
\providecommand \bibfnamefont [1]{#1}%
\providecommand \citenamefont [1]{#1}%
\providecommand \href@noop [0]{\@secondoftwo}%
\providecommand \href [0]{\begingroup \@sanitize@url \@href}%
\providecommand \@href[1]{\@@startlink{#1}\@@href}%
\providecommand \@@href[1]{\endgroup#1\@@endlink}%
\providecommand \@sanitize@url [0]{\catcode `\\12\catcode `\$12\catcode
  `\&12\catcode `\#12\catcode `\^12\catcode `\_12\catcode `\%12\relax}%
\providecommand \@@startlink[1]{}%
\providecommand \@@endlink[0]{}%
\providecommand \url  [0]{\begingroup\@sanitize@url \@url }%
\providecommand \@url [1]{\endgroup\@href {#1}{\urlprefix }}%
\providecommand \urlprefix  [0]{URL }%
\providecommand \Eprint [0]{\href }%
\@ifxundefined \urlstyle {%
  \providecommand \doi  [0]{\begingroup \@sanitize@url \@doi}%
  \providecommand \@doi [1]{\endgroup \@@startlink {\doibase
  #1}doi:\discretionary {}{}{}#1\@@endlink }%
}{%
  \providecommand \doi  [0]{doi:\discretionary{}{}{}\begingroup
  \urlstyle{rm}\Url }%
}%
\providecommand \doibase [0]{http://dx.doi.org/}%
\providecommand \Doi [0]{\begingroup \@sanitize@url \@Doi }%
\providecommand \@Doi  [1]{\endgroup\@@startlink{\doibase#1}\@@Doi}%
\providecommand \@@Doi [1]{#1\@@endlink}%
\providecommand \selectlanguage [0]{\@gobble}%
\providecommand \bibinfo  [0]{\@secondoftwo}%
\providecommand \bibfield  [0]{\@secondoftwo}%
\providecommand \translation [1]{[#1]}%
\providecommand \BibitemOpen [0]{}%
\providecommand \bibitemStop [0]{}%
\providecommand \bibitemNoStop [0]{.\EOS\space}%
\providecommand \EOS [0]{\spacefactor3000\relax}%
\providecommand \BibitemShut  [1]{\csname bibitem#1\endcsname}%
\bibitem [{\citenamefont {Dicke}(1954)}]{dicke54}%
  \BibitemOpen
  \bibfield  {author} {\bibinfo {author} {\bibfnamefont {R.~H.}\ \bibnamefont
  {Dicke}},\ }\Doi {10.1103/PhysRev.93.99} {\bibfield  {journal} {\bibinfo
  {journal} {Phys. Rev.},\ }\textbf {\bibinfo {volume} {93}},\ \bibinfo {pages}
  {99} (\bibinfo {year} {1954})}\BibitemShut {NoStop}%
\bibitem [{\citenamefont {Gross}\ and\ \citenamefont
  {Haroche}(1982)}]{gross82}%
  \BibitemOpen
  \bibfield  {author} {\bibinfo {author} {\bibfnamefont {M.}~\bibnamefont
  {Gross}}\ and\ \bibinfo {author} {\bibfnamefont {S.}~\bibnamefont
  {Haroche}},\ }\Doi {10.1016/0370-1573(82)90102-8} {\bibfield  {journal}
  {\bibinfo  {journal} {Physics Reports},\ }\textbf {\bibinfo {volume} {93}},\
  \bibinfo {pages} {301} (\bibinfo {year} {1982})}\BibitemShut {NoStop}%
\bibitem [{\citenamefont {DeVoe}\ and\ \citenamefont {Brewer}(1996)}]{devoe96}%
  \BibitemOpen
  \bibfield  {author} {\bibinfo {author} {\bibfnamefont {R.~G.}\ \bibnamefont
  {DeVoe}}\ and\ \bibinfo {author} {\bibfnamefont {R.~G.}\ \bibnamefont
  {Brewer}},\ }\Doi {10.1103/PhysRevLett.76.2049} {\bibfield  {journal}
  {\bibinfo  {journal} {Phys. Rev. Lett.},\ }\textbf {\bibinfo {volume} {76}},\
  \bibinfo {pages} {2049} (\bibinfo {year} {1996})}\BibitemShut {NoStop}%
\bibitem [{\citenamefont {Hettich}\ \emph {et~al.}(2002)\citenamefont
  {Hettich}, \citenamefont {Schmitt}, \citenamefont {Zitzmann}, \citenamefont
  {K\"{u}hn}, \citenamefont {Gerhardt},\ and\ \citenamefont
  {Sandoghdar}}]{hettich02}%
  \BibitemOpen
  \bibfield  {author} {\bibinfo {author} {\bibfnamefont {C.}~\bibnamefont
  {Hettich}}, \bibinfo {author} {\bibfnamefont {C.}~\bibnamefont {Schmitt}},
  \bibinfo {author} {\bibfnamefont {J.}~\bibnamefont {Zitzmann}}, \bibinfo
  {author} {\bibfnamefont {S.}~\bibnamefont {K\"{u}hn}}, \bibinfo {author}
  {\bibfnamefont {I.}~\bibnamefont {Gerhardt}}, \ and\ \bibinfo {author}
  {\bibfnamefont {V.}~\bibnamefont {Sandoghdar}},\ }\Doi
  {10.1126/science.1075606} {\bibfield  {journal} {\bibinfo  {journal}
  {Science},\ }\textbf {\bibinfo {volume} {298}},\ \bibinfo {pages} {385}
  (\bibinfo {year} {2002})}\BibitemShut {NoStop}%
\bibitem [{\citenamefont {Scheibner}\ \emph {et~al.}(2007)\citenamefont
  {Scheibner}, \citenamefont {Schmidt}, \citenamefont {Worschech},
  \citenamefont {Forchel}, \citenamefont {Bacher}, \citenamefont {Passow},\
  and\ \citenamefont {Hommel}}]{scheibner07}%
  \BibitemOpen
  \bibfield  {author} {\bibinfo {author} {\bibfnamefont {M.}~\bibnamefont
  {Scheibner}}, \bibinfo {author} {\bibfnamefont {T.}~\bibnamefont {Schmidt}},
  \bibinfo {author} {\bibfnamefont {L.}~\bibnamefont {Worschech}}, \bibinfo
  {author} {\bibfnamefont {A.}~\bibnamefont {Forchel}}, \bibinfo {author}
  {\bibfnamefont {G.}~\bibnamefont {Bacher}}, \bibinfo {author} {\bibfnamefont
  {T.}~\bibnamefont {Passow}}, \ and\ \bibinfo {author} {\bibfnamefont
  {D.}~\bibnamefont {Hommel}},\ }\Doi {10.1038/nphys494} {\bibfield  {journal}
  {\bibinfo  {journal} {Nature Phys.},\ }\textbf {\bibinfo {volume} {3}},\
  \bibinfo {pages} {106} (\bibinfo {year} {2007})}\BibitemShut {NoStop}%
\bibitem [{\citenamefont {Agarwal}(1974)}]{agarwal74}%
  \BibitemOpen
  \bibfield  {author} {\bibinfo {author} {\bibfnamefont {G.~S.}\ \bibnamefont
  {Agarwal}},\ }\Doi {10.1007/BFb0042382} {\bibfield  {journal} {\bibinfo
  {journal} {{Springer Tracts in Modern Physics}},\ }\textbf {\bibinfo {volume}
  {70}},\ \bibinfo {pages} {1} (\bibinfo {year} {1974})}\BibitemShut {NoStop}%
\bibitem [{\citenamefont {Beige}\ and\ \citenamefont
  {Hegerfeldt}(1998)}]{beige98}%
  \BibitemOpen
  \bibfield  {author} {\bibinfo {author} {\bibfnamefont {A.}~\bibnamefont
  {Beige}}\ and\ \bibinfo {author} {\bibfnamefont {G.~C.}\ \bibnamefont
  {Hegerfeldt}},\ }\Doi {10.1103/PhysRevA.58.4133} {\bibfield  {journal}
  {\bibinfo  {journal} {Phys. Rev. A},\ }\textbf {\bibinfo {volume} {58}},\
  \bibinfo {pages} {4133} (\bibinfo {year} {1998})}\BibitemShut {NoStop}%
\bibitem [{\citenamefont {Boller}\ \emph {et~al.}(1991)\citenamefont {Boller},
  \citenamefont {{Imamo\u{g}lu}},\ and\ \citenamefont {Harris}}]{boller91}%
  \BibitemOpen
  \bibfield  {author} {\bibinfo {author} {\bibfnamefont {K.-J.}\ \bibnamefont
  {Boller}}, \bibinfo {author} {\bibfnamefont {A.}~\bibnamefont
  {{Imamo\u{g}lu}}}, \ and\ \bibinfo {author} {\bibfnamefont {S.~E.}\
  \bibnamefont {Harris}},\ }\Doi {10.1103/PhysRevLett.66.2593} {\bibfield
  {journal} {\bibinfo  {journal} {Phys. Rev. Lett.},\ }\textbf {\bibinfo
  {volume} {66}},\ \bibinfo {pages} {2593} (\bibinfo {year}
  {1991})}\BibitemShut {NoStop}%
\bibitem [{\citenamefont {Mohapatra}\ \emph {et~al.}(2007)\citenamefont
  {Mohapatra}, \citenamefont {Jackson},\ and\ \citenamefont
  {Adams}}]{mohapatra07}%
  \BibitemOpen
  \bibfield  {author} {\bibinfo {author} {\bibfnamefont {A.~K.}\ \bibnamefont
  {Mohapatra}}, \bibinfo {author} {\bibfnamefont {T.~R.}\ \bibnamefont
  {Jackson}}, \ and\ \bibinfo {author} {\bibfnamefont {C.~S.}\ \bibnamefont
  {Adams}},\ }\Doi {10.1103/PhysRevLett.98.113003} {\bibfield  {journal}
  {\bibinfo  {journal} {Phys. Rev. Lett.},\ }\textbf {\bibinfo {volume} {98}},\
  \bibinfo {pages} {113003} (\bibinfo {year} {2007})}\BibitemShut {NoStop}%
\bibitem [{\citenamefont {Pritchard}\ \emph {et~al.}(2010)\citenamefont
  {Pritchard}, \citenamefont {Maxwell}, \citenamefont {Gauguet}, \citenamefont
  {Weatherill}, \citenamefont {Jones},\ and\ \citenamefont
  {Adams}}]{pritchard10}%
  \BibitemOpen
  \bibfield  {author} {\bibinfo {author} {\bibfnamefont {J.~D.}\ \bibnamefont
  {Pritchard}}, \bibinfo {author} {\bibfnamefont {D.}~\bibnamefont {Maxwell}},
  \bibinfo {author} {\bibfnamefont {A.}~\bibnamefont {Gauguet}}, \bibinfo
  {author} {\bibfnamefont {K.~J.}\ \bibnamefont {Weatherill}}, \bibinfo
  {author} {\bibfnamefont {M.~P.~A.}\ \bibnamefont {Jones}}, \ and\ \bibinfo
  {author} {\bibfnamefont {C.~S.}\ \bibnamefont {Adams}},\ }\Doi
  {10.1103/PhysRevLett.105.193603} {\bibfield  {journal} {\bibinfo  {journal}
  {Phys. Rev. Lett.},\ }\textbf {\bibinfo {volume} {105}},\ \bibinfo {pages}
  {193603} (\bibinfo {year} {2010})}\BibitemShut {NoStop}%
\bibitem [{\citenamefont {Lukin}\ \emph {et~al.}(2001)\citenamefont {Lukin},
  \citenamefont {Fleischhauer}, \citenamefont {Cote}, \citenamefont {Duan},
  \citenamefont {Jaksch}, \citenamefont {Cirac},\ and\ \citenamefont
  {Zoller}}]{lukin01}%
  \BibitemOpen
  \bibfield  {author} {\bibinfo {author} {\bibfnamefont {M.~D.}\ \bibnamefont
  {Lukin}}, \bibinfo {author} {\bibfnamefont {M.}~\bibnamefont {Fleischhauer}},
  \bibinfo {author} {\bibfnamefont {R.}~\bibnamefont {Cote}}, \bibinfo {author}
  {\bibfnamefont {L.~M.}\ \bibnamefont {Duan}}, \bibinfo {author}
  {\bibfnamefont {D.}~\bibnamefont {Jaksch}}, \bibinfo {author} {\bibfnamefont
  {J.~I.}\ \bibnamefont {Cirac}}, \ and\ \bibinfo {author} {\bibfnamefont
  {P.}~\bibnamefont {Zoller}},\ }\Doi {10.1103/PhysRevLett.87.037901}
  {\bibfield  {journal} {\bibinfo  {journal} {Phys. Rev. Lett.},\ }\textbf
  {\bibinfo {volume} {87}},\ \bibinfo {pages} {037901} (\bibinfo {year}
  {2001})}\BibitemShut {NoStop}%
\bibitem [{\citenamefont {Saffman}\ \emph {et~al.}(2010)\citenamefont
  {Saffman}, \citenamefont {Walker},\ and\ \citenamefont
  {M\o{}lmer}}]{saffman10}%
  \BibitemOpen
  \bibfield  {author} {\bibinfo {author} {\bibfnamefont {M.}~\bibnamefont
  {Saffman}}, \bibinfo {author} {\bibfnamefont {T.~G.}\ \bibnamefont {Walker}},
  \ and\ \bibinfo {author} {\bibfnamefont {K.}~\bibnamefont {M\o{}lmer}},\
  }\Doi {10.1103/RevModPhys.82.2313} {\bibfield  {journal} {\bibinfo  {journal}
  {Rev. Mod. Phys.},\ }\textbf {\bibinfo {volume} {82}},\ \bibinfo {pages}
  {2313} (\bibinfo {year} {2010})}\BibitemShut {NoStop}%
\bibitem [{\citenamefont {Ates}\ \emph {et~al.}(2011)\citenamefont {Ates},
  \citenamefont {Sevin\c{c}li},\ and\ \citenamefont {Pohl}}]{ates11}%
  \BibitemOpen
  \bibfield  {author} {\bibinfo {author} {\bibfnamefont {C.}~\bibnamefont
  {Ates}}, \bibinfo {author} {\bibfnamefont {S.}~\bibnamefont {Sevin\c{c}li}},
  \ and\ \bibinfo {author} {\bibfnamefont {T.}~\bibnamefont {Pohl}},\ }\Doi
  {10.1103/PhysRevA.83.041802} {\bibfield  {journal} {\bibinfo  {journal}
  {Phys. Rev. A},\ }\textbf {\bibinfo {volume} {83}},\ \bibinfo {pages}
  {041802} (\bibinfo {year} {2011})}\BibitemShut {NoStop}%
\bibitem [{\citenamefont {Sevin\c{c}li}\ \emph {et~al.}(2011)\citenamefont
  {Sevin\c{c}li}, \citenamefont {Henkel}, \citenamefont {Ates},\ and\
  \citenamefont {Pohl}}]{sevincli11}%
  \BibitemOpen
  \bibfield  {author} {\bibinfo {author} {\bibfnamefont {S.}~\bibnamefont
  {Sevin\c{c}li}}, \bibinfo {author} {\bibfnamefont {N.}~\bibnamefont
  {Henkel}}, \bibinfo {author} {\bibfnamefont {C.}~\bibnamefont {Ates}}, \ and\
  \bibinfo {author} {\bibfnamefont {T.}~\bibnamefont {Pohl}},\ }\href@noop {}
  {\enquote {\bibinfo {title} {{Nonlocal Nonlinear Optics in cold Rydberg
  Gases}},}\ } (\bibinfo {year} {2011}),\ \Eprint
  {http://arxiv.org/abs/1106.2001} {arXiv:1106.2001 [atom-ph]} \BibitemShut
  {NoStop}%
\bibitem [{\citenamefont {Honer}\ \emph {et~al.}(2011)\citenamefont {Honer},
  \citenamefont {Weimer}, \citenamefont {B\"{u}chler}, \citenamefont
  {L\"{o}w},\ and\ \citenamefont {Pfau}}]{honer11}%
  \BibitemOpen
  \bibfield  {author} {\bibinfo {author} {\bibfnamefont {J.}~\bibnamefont
  {Honer}}, \bibinfo {author} {\bibfnamefont {H.}~\bibnamefont {Weimer}},
  \bibinfo {author} {\bibfnamefont {H.~P.}\ \bibnamefont {B\"{u}chler}},
  \bibinfo {author} {\bibfnamefont {R.}~\bibnamefont {L\"{o}w}}, \ and\
  \bibinfo {author} {\bibfnamefont {T.}~\bibnamefont {Pfau}},\ }\href@noop {}
  {\enquote {\bibinfo {title} {{Artificial atoms can do more than atoms:
  Deterministic single photon subtraction from arbitrary light fields}},}\ }
  (\bibinfo {year} {2011}),\ \Eprint {http://arxiv.org/abs/1103.1319}
  {arXiv:1103.1319 [quant-ph]} \BibitemShut {NoStop}%
\bibitem [{\citenamefont {Shahmoon}\ \emph {et~al.}(2011)\citenamefont
  {Shahmoon}, \citenamefont {Kurizki}, \citenamefont {Fleischhauer},\ and\
  \citenamefont {Petrosyan}}]{shahmoon10}%
  \BibitemOpen
  \bibfield  {author} {\bibinfo {author} {\bibfnamefont {E.}~\bibnamefont
  {Shahmoon}}, \bibinfo {author} {\bibfnamefont {G.}~\bibnamefont {Kurizki}},
  \bibinfo {author} {\bibfnamefont {M.}~\bibnamefont {Fleischhauer}}, \ and\
  \bibinfo {author} {\bibfnamefont {D.}~\bibnamefont {Petrosyan}},\ }\Doi
  {10.1103/PhysRevA.83.033806} {\bibfield  {journal} {\bibinfo  {journal}
  {Phys. Rev. A},\ }\textbf {\bibinfo {volume} {83}},\ \bibinfo {pages}
  {033806} (\bibinfo {year} {2011})}\BibitemShut {NoStop}%
\bibitem [{\citenamefont {Gorshkov}\ \emph {et~al.}(2011)\citenamefont
  {Gorshkov}, \citenamefont {Otterback}, \citenamefont {Fleischhauer},
  \citenamefont {Pohl},\ and\ \citenamefont {Lukin}}]{gorshkov11}%
  \BibitemOpen
  \bibfield  {author} {\bibinfo {author} {\bibfnamefont {A.~V.}\ \bibnamefont
  {Gorshkov}}, \bibinfo {author} {\bibfnamefont {J.}~\bibnamefont {Otterback}},
  \bibinfo {author} {\bibfnamefont {M.}~\bibnamefont {Fleischhauer}}, \bibinfo
  {author} {\bibfnamefont {T.}~\bibnamefont {Pohl}}, \ and\ \bibinfo {author}
  {\bibfnamefont {M.~D.}\ \bibnamefont {Lukin}},\ }\href@noop {} {\enquote
  {\bibinfo {title} {{Photon-Photon Interactions via Rydberg Blockage}},}\ }
  (\bibinfo {year} {2011}),\ \Eprint {http://arxiv.org/abs/1103.3700}
  {arXiv:1103.3700 [quant-ph]} \BibitemShut {NoStop}%
\bibitem [{\citenamefont {Petrosyan}\ \emph {et~al.}(2011)\citenamefont
  {Petrosyan}, \citenamefont {Otterbach},\ and\ \citenamefont
  {Fleischhauer}}]{petrosyan11}%
  \BibitemOpen
  \bibfield  {author} {\bibinfo {author} {\bibfnamefont {D.}~\bibnamefont
  {Petrosyan}}, \bibinfo {author} {\bibfnamefont {J.}~\bibnamefont
  {Otterbach}}, \ and\ \bibinfo {author} {\bibfnamefont {M.}~\bibnamefont
  {Fleischhauer}},\ }\href@noop {} {\enquote {\bibinfo {title}
  {{Electromagnetically Induced Transparency with Rydberg Atoms}},}\ }
  (\bibinfo {year} {2011}),\ \Eprint {http://arxiv.org/abs/1106.1360}
  {arXiv:1106.1360 [quant-ph]} \BibitemShut {NoStop}%
\bibitem [{\citenamefont {Olmos}\ and\ \citenamefont
  {Lesanovsky}(2010)}]{olmos10}%
  \BibitemOpen
  \bibfield  {author} {\bibinfo {author} {\bibfnamefont {B.}~\bibnamefont
  {Olmos}}\ and\ \bibinfo {author} {\bibfnamefont {I.}~\bibnamefont
  {Lesanovsky}},\ }\Doi {10.1103/PhysRevA.82.063404} {\bibfield  {journal}
  {\bibinfo  {journal} {Phys. Rev. A},\ }\textbf {\bibinfo {volume} {82}},\
  \bibinfo {pages} {063404} (\bibinfo {year} {2010})}\BibitemShut {NoStop}%
\bibitem [{\citenamefont {Pohl}\ \emph {et~al.}(2010)\citenamefont {Pohl},
  \citenamefont {Demler},\ and\ \citenamefont {Lukin}}]{pohl10}%
  \BibitemOpen
  \bibfield  {author} {\bibinfo {author} {\bibfnamefont {T.}~\bibnamefont
  {Pohl}}, \bibinfo {author} {\bibfnamefont {E.}~\bibnamefont {Demler}}, \ and\
  \bibinfo {author} {\bibfnamefont {M.~D.}\ \bibnamefont {Lukin}},\ }\Doi
  {10.1103/PhysRevLett.104.043002} {\bibfield  {journal} {\bibinfo  {journal}
  {Phys. Rev. Lett.},\ }\textbf {\bibinfo {volume} {104}},\ \bibinfo {pages}
  {043002} (\bibinfo {year} {2010})}\BibitemShut {NoStop}%
\bibitem [{\citenamefont {Nielsen}\ and\ \citenamefont
  {M{\o}lmer}(2010)}]{nielsen10}%
  \BibitemOpen
  \bibfield  {author} {\bibinfo {author} {\bibfnamefont {A.~E.~B.}\
  \bibnamefont {Nielsen}}\ and\ \bibinfo {author} {\bibfnamefont
  {K.}~\bibnamefont {M{\o}lmer}},\ }\Doi {10.1103/PhysRevA.81.043822}
  {\bibfield  {journal} {\bibinfo  {journal} {Phys. Rev. A},\ }\textbf
  {\bibinfo {volume} {81}},\ \bibinfo {pages} {043822} (\bibinfo {year}
  {2010})}\BibitemShut {NoStop}%
\bibitem [{\citenamefont {Wilk}\ \emph {et~al.}(2010)\citenamefont {Wilk},
  \citenamefont {Ga\"etan}, \citenamefont {Evellin}, \citenamefont {Wolters},
  \citenamefont {Miroshnychenko}, \citenamefont {Grangier},\ and\ \citenamefont
  {Browaeys}}]{wilk10}%
  \BibitemOpen
  \bibfield  {author} {\bibinfo {author} {\bibfnamefont {T.}~\bibnamefont
  {Wilk}}, \bibinfo {author} {\bibfnamefont {A.}~\bibnamefont {Ga\"etan}},
  \bibinfo {author} {\bibfnamefont {C.}~\bibnamefont {Evellin}}, \bibinfo
  {author} {\bibfnamefont {J.}~\bibnamefont {Wolters}}, \bibinfo {author}
  {\bibfnamefont {Y.}~\bibnamefont {Miroshnychenko}}, \bibinfo {author}
  {\bibfnamefont {P.}~\bibnamefont {Grangier}}, \ and\ \bibinfo {author}
  {\bibfnamefont {A.}~\bibnamefont {Browaeys}},\ }\Doi
  {10.1103/PhysRevLett.104.010502} {\bibfield  {journal} {\bibinfo  {journal}
  {Phys. Rev. Lett.},\ }\textbf {\bibinfo {volume} {104}},\ \bibinfo {pages}
  {010502} (\bibinfo {year} {2010})}\BibitemShut {NoStop}%
\bibitem [{\citenamefont {Isenhower}\ \emph {et~al.}(2010)\citenamefont
  {Isenhower}, \citenamefont {Urban}, \citenamefont {Zhang}, \citenamefont
  {Gill}, \citenamefont {Henage}, \citenamefont {Johnson}, \citenamefont
  {Walker},\ and\ \citenamefont {Saffman}}]{isenhower10}%
  \BibitemOpen
  \bibfield  {author} {\bibinfo {author} {\bibfnamefont {L.}~\bibnamefont
  {Isenhower}}, \bibinfo {author} {\bibfnamefont {E.}~\bibnamefont {Urban}},
  \bibinfo {author} {\bibfnamefont {X.~L.}\ \bibnamefont {Zhang}}, \bibinfo
  {author} {\bibfnamefont {A.~T.}\ \bibnamefont {Gill}}, \bibinfo {author}
  {\bibfnamefont {T.}~\bibnamefont {Henage}}, \bibinfo {author} {\bibfnamefont
  {T.~A.}\ \bibnamefont {Johnson}}, \bibinfo {author} {\bibfnamefont {T.~G.}\
  \bibnamefont {Walker}}, \ and\ \bibinfo {author} {\bibfnamefont
  {M.}~\bibnamefont {Saffman}},\ }\Doi {10.1103/PhysRevLett.104.010503}
  {\bibfield  {journal} {\bibinfo  {journal} {Phys. Rev. Lett.},\ }\textbf
  {\bibinfo {volume} {104}},\ \bibinfo {pages} {010503} (\bibinfo {year}
  {2010})}\BibitemShut {NoStop}%
\bibitem [{\citenamefont {Loudon}(1997)}]{loudon97}%
  \BibitemOpen
  \bibfield  {author} {\bibinfo {author} {\bibfnamefont {R.}~\bibnamefont
  {Loudon}},\ }\href@noop {} {\emph {\bibinfo {title} {The Quantum Theory of
  Light}}},\ \bibinfo {edition} {2nd}\ ed.\ (\bibinfo  {publisher} {OUP, UK},\
  \bibinfo {year} {1997})\BibitemShut {NoStop}%
\bibitem [{\citenamefont {M{\o}lmer}\ and\ \citenamefont
  {Castin}(1996)}]{molmer96}%
  \BibitemOpen
  \bibfield  {author} {\bibinfo {author} {\bibfnamefont {K.}~\bibnamefont
  {M{\o}lmer}}\ and\ \bibinfo {author} {\bibfnamefont {Y.}~\bibnamefont
  {Castin}},\ }\Doi {10.1088/1355-5111/8/1/007} {\bibfield  {journal} {\bibinfo
   {journal} {Quantum Semiclass. Opt.},\ }\textbf {\bibinfo {volume} {8}},\
  \bibinfo {pages} {49} (\bibinfo {year} {1996})}\BibitemShut {NoStop}%
\bibitem [{\citenamefont {Lindblad}(1976)}]{lindblad76}%
  \BibitemOpen
  \bibfield  {author} {\bibinfo {author} {\bibfnamefont {G.}~\bibnamefont
  {Lindblad}},\ }\Doi {10.1007/BF01608499} {\bibfield  {journal} {\bibinfo
  {journal} {Comm. Math. Phys.},\ }\textbf {\bibinfo {volume} {48}},\ \bibinfo
  {pages} {119} (\bibinfo {year} {1976})}\BibitemShut {NoStop}%
\bibitem [{\citenamefont {Kimble}\ and\ \citenamefont
  {Mandel}(1976)}]{kimble76}%
  \BibitemOpen
  \bibfield  {author} {\bibinfo {author} {\bibfnamefont {H.~J.}\ \bibnamefont
  {Kimble}}\ and\ \bibinfo {author} {\bibfnamefont {L.}~\bibnamefont
  {Mandel}},\ }\Doi {10.1103/PhysRevA.13.2123} {\bibfield  {journal} {\bibinfo
  {journal} {Phys. Rev. A},\ }\textbf {\bibinfo {volume} {13}},\ \bibinfo
  {pages} {2123} (\bibinfo {year} {1976})}\BibitemShut {NoStop}%
\bibitem [{\citenamefont {Kimble}\ \emph {et~al.}(1977)\citenamefont {Kimble},
  \citenamefont {Dagenais},\ and\ \citenamefont {Mandel}}]{kimble77}%
  \BibitemOpen
  \bibfield  {author} {\bibinfo {author} {\bibfnamefont {H.~J.}\ \bibnamefont
  {Kimble}}, \bibinfo {author} {\bibfnamefont {M.}~\bibnamefont {Dagenais}}, \
  and\ \bibinfo {author} {\bibfnamefont {L.}~\bibnamefont {Mandel}},\ }\Doi
  {10.1103/PhysRevLett.39.691} {\bibfield  {journal} {\bibinfo  {journal}
  {Phys. Rev. Lett.},\ }\textbf {\bibinfo {volume} {39}},\ \bibinfo {pages}
  {691} (\bibinfo {year} {1977})}\BibitemShut {NoStop}%
\bibitem [{\citenamefont {M\o{}ller}\ \emph {et~al.}(2008)\citenamefont
  {M\o{}ller}, \citenamefont {Madsen},\ and\ \citenamefont
  {M\o{}lmer}}]{moller08}%
  \BibitemOpen
  \bibfield  {author} {\bibinfo {author} {\bibfnamefont {D.}~\bibnamefont
  {M\o{}ller}}, \bibinfo {author} {\bibfnamefont {L.~B.}\ \bibnamefont
  {Madsen}}, \ and\ \bibinfo {author} {\bibfnamefont {K.}~\bibnamefont
  {M\o{}lmer}},\ }\Doi {10.1103/PhysRevLett.100.170504} {\bibfield  {journal}
  {\bibinfo  {journal} {Phys. Rev. Lett.},\ }\textbf {\bibinfo {volume}
  {100}},\ \bibinfo {eid} {170504} (\bibinfo {year} {2008})}\BibitemShut
  {NoStop}%
\end{thebibliography}
\end{document}